\documentclass[preprint,review]{elsarticle}

\usepackage{amsmath}
\usepackage{amsfonts}
\usepackage{array}
\usepackage{siunitx}
\usepackage{mathtools}
\usepackage{tablefootnote}

\usepackage{lineno}

\modulolinenumbers[5]

\journal{}

\begin{document}

\begin{frontmatter}

\title{Dedicated beam position monitor pair for model-independent lattice characterization at NSLS-II}

\author{Yongjun Li\corref{correspondingauthor}}
\cortext[correspondingauthor]{Corresponding author}
\ead{yli@bnl.gov}
\author{Kiman Ha}
\author{Danny Padrazo} 
\author{Bernard Kosciuk} 
\author{Belkacem Bacha} 
\author{Michael Seegitz}
\author{Robert Rainer}
\author{Joseph Mead}
\author{Xi Yang}
\author{Yuke Tian}
\author{Robert Todd}
\author{Victor Smaluk}
\address{Brookhaven National Laboratory, Upton, NY, 11973, USA}

\author{Weixing Cheng}
\address{Argonne National Laboratory, Lemont, IL, 60439, USA}

\begin{abstract}
 This paper reports recent lattice characterization results obtained at the National Synchrotron Light Source II (NSLS-II) storage ring, conducted without reliance on a lattice model. A pair of beam position monitors (BPMs) with bunch-by-bunch (B$\times$B) resolution, were recently installed in a section of the storage ring free of magnetic fields. The new BPM pair measured the beam, or bunch's transverse Poincar\'e map precisely after the beam was excited. Linear one-turn-matrices (OTM) were then derived, and from these, the 4-dimensional coupled Twiss parameters were extracted at the locations of the BPM pair. By normalizing beam oscillation amplitudes with the Twiss parameters, the global action-variables were obtained. These action-variables facilitated the measurement of the local Twiss parameters observed by other BPMs independent on lattice model. This method is general, and particularly useful in certain scenarios such as a round beam mode in a diffraction-limited light source ring. We applied it to assess both weakly and strongly coupled lattices at the NSLS-II ring. Through analysis of the strongly coupled lattice, the quadrupole tilt errors were estimated to be less than 400 \si{\mu}rad.  Utilizing the BPMs' B$\times$B resolution, for the first time we observed the variations of the linear lattice along a long bunch-train.
\end{abstract}


\end{frontmatter}


\section{\label{sect:intro}Introduction}
 Model-independent lattice characterization methods using multi-turn data in storage rings are well-documented across numerous studies, including ref.~\cite{irwin1999,wang2004untangling,huang2010lattice,riemann2011model}. This method is particularly useful when precise lattice models are unavailable. For example, some extremely low-emittance light source rings will operate in a round beam mode with on-resonance tunes to extend beam lifetime, as noted by~\cite{Borland2017,Steier2018}. In such cases, strong linear coupling can result from unidentified random errors. Another scenario involves the obvious distortion of the design lattice due to magnet imperfections and/or the influence of insertion devices (IDs). Although ID fields can be taken into account at the lattice design stage, during routine operations, their actual field can dynamically vary due to changes in their gaps, phase adjustments and compensation coils. Therefore, design models might not be able to accurately represent a machine's real-world operation. Given that at large user facilities, their storage rings often host several tens of IDs, a model-independent lattice characterization might be considered under these conditions.

 A direct method to characterize the linear lattice involves reconstructing the one-turn-matrix (OTM) using multi-turn data captured by BPMs~\cite{fischer2003}. Assuming the transfer matrix between a pair of the BPMs is known, with consecutive turn-by-turn (T$\times$T) data, the Poincar\'e map can be constructed to fit OTMs. However, this assumption is only approximately valid due to unidentified random errors in the presence of real magnets, except when the BPM pair is inside field-free region. In field-free regions the transverse momenta (trajectory slopes) of the beam can be precisely determined using the formula:
 \begin{equation}\label{eq:slope}
    p_x\approx x^{\prime}=\frac{x_2-x_1}{L},
 \end{equation}
 where $x_{1,2}$ are the readings from the upstream and downstream BPMs at the same turn, and $L$ is the distance between them. With well-calibrated and aligned BPMs, the projected trajectories in the phase space can be accurately reconstructed.  To exploit this capability, a BPM pair was recently installed in the cell 1 straight section of the NSLS-II ring. This section is reserved for beam diagnostics and will remain free of magnets for the foreseeable future.
 
 In what follows, we summarize some recent lattice measurement results utilizing the new BPM pair: Sect.~\ref{sect:BPM} introduces the new BPMs' electronics and compares their T$\times$T resolution performance with other regular BPMs. In Sect.~\ref{sect:coupledOptics}, the lattice characterization results for both weak and strongly coupled lattices are presented. Through fitting the measurement results, the NSLS-II ring's quadrupole tilt errors were estimated, for the first time, based on a beam test. A preliminary observation of the non-linearity was also presented. Sect.~\ref{sect:BxB} reports the measurements of each individual bunch's Twiss parameters in a long bunch-train using their bunch-by-bunch (B$\times$B) resolution. A brief summary is given in Sect.~\ref{sect:summary}.

\section{\label{sect:BPM}BPMs and their electronics}
\begin{figure}[ht]
    \centering
    \includegraphics[width=\columnwidth]{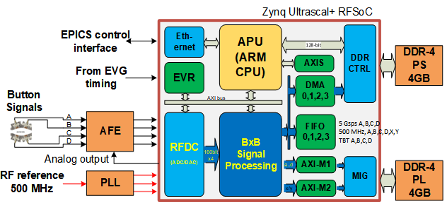}
    \includegraphics[width=\columnwidth]{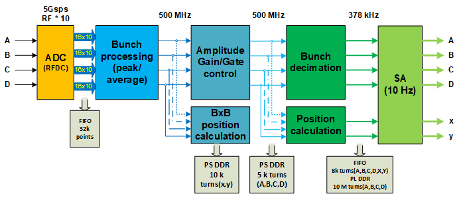}
    \caption{(Colored) Schematic diagrams of the B$\times$B BPM electronics. Top: Diagram of the FPGA firmware functionalities. Bottom: Diagram of the FPGA signal processing.}
    \label{fig:BPMfunsignal}
 \end{figure}

 The recently installed BPM pair is situated within the cell 1 straight section, separated by 5.14 \si{m}. Their transverse cross-section geometry is the same as our existing regular BPMs as detailed by ~\cite{singh2013nsls}. Besides regular functionalities, these BPMs feature advanced electronics capable of providing B$\times$B resolution, allowing detailed exploration of the beam dynamics at the bunch level. The electronics of these BPMs utilize a Radio Frequency System-on-Chip (RFSoC) Field Programmable Gate Array (FPGA). The RFSoC FPGA includes eight 14-bit Analog-to-Digital Converters (ADCs) with 5 Giga Samples Per Second (GSPS), directly sampling the BPMs' four-button signals. The BPM signals are synchronized with the 500 MHz RF reference clock to align to each bunch. The system's chassis incorporates a Xilinx ZCU208 FPGA evaluation board and an RF breakout Balun board to interface with the single-end analog input/output signals. Additionally, a custom-designed four-channel RF switchboard and a hybrid Analog Front-End (AFE) module are employed for beam signal conditioning. A linear AC/DC power supply minimizes amplifier noise and an external Double Data Rate $4^{th}$-generation (DDR4) memory block is integrated for fast, extensive data capture. Beam data is relayed to Input-Output Controllers (IOCs) and client computers via the Experimental Physics and Industrial Control System (EPICS). The functionalities of the electronics firmware and signal processing are illustrated in Fig.~\ref{fig:BPMfunsignal}. Further details on the BPM electronics are discussed in ref.~\cite{ha2024}.

\begin{figure}[ht]
    \centering    \includegraphics[width=\columnwidth]{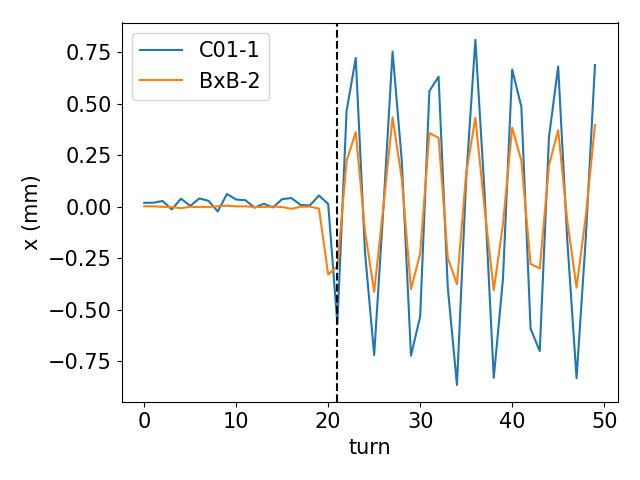}
    \caption{(Colored) Comparison of the T$\times$T readings of a new B$\times$B BPM (B$\times$B-2) and its nearest regular counterpart, BPM (C01-1), when the beam was excited at the $21^{st}$ turn. The new BPM's noise fluctuation is significantly smaller than the regular BPM before beam excitation.}
    \label{fig:betterBPM}
 \end{figure}

 The performance of the new BPMs' T$\times$T data acquisition was compared to regular BPMs under identical conditions. T$\times$T data from a new BPM (B$\times$B-2) and its nearest regular counterpart (BPM C01-1) are shown in Fig.~\ref{fig:betterBPM}. The T$\times$T data prior to the $21^{st}$ turn, when the beam was not kicked and was expected to remain stable. In this plot, the new BPM exhibited significantly less fluctuation ($\approx4\,\si{\mu m}$) than its regular counterpart ($\approx22\,\si{\mu m}$). In principle, T$\times$T fluctuations can generally be smoothed out by averaging over multiple turns for either the slow (10 Hz) or fast (10 kHz) orbit control. This average results in a minimized impact on routine machine operations. The new electronics, however, are particularly advantageous when precise T$\times$T data is needed to measure the lattice functions. Additionally, the new BPM pair's B$\times$B resolution is approximately 2 to 3 times less precise than the regular BPM's T$\times$T resolution, and will be discussed in Sect.~\ref{sect:BxB}.
 
\section{\label{sect:coupledOptics}Measurement of generally coupled linear lattices}
\subsection{Introduction to method}
 In a generally coupled linear lattice, beam Betatron oscillation is observable by BPMs following excitation with the pulse kickers, referred to as pingers. The turn-by-turn beam trajectory, as read by a BPM, can be mathematically represented as:
 \begin{equation}\label{eq:coupledTbT}
  \begin{bmatrix}
  x_i\\ y_i
  \end{bmatrix}
  =
  \begin{bmatrix}
  A_{1,x}\cos(i\cdot2\pi\nu_1+\phi_{1,x})+
  A_{2,x}\cos(i\cdot2\pi\nu_2+\phi_{2,x})+x_{co}\\
  A_{1,y}\cos(i\cdot2\pi\nu_1+\phi_{1,y})+
  A_{2,y}\cos(i\cdot2\pi\nu_2+\phi_{2,y})+y_{co}
  \end{bmatrix},
 \end{equation}
 where $\nu_{1,2}$ are the tunes of two oscillation modes, $i$ is the index of turns, $\phi_{(1,2),{(x,y)}}$ are the initial phases, and $x_{co}$ and $y_{co}$ represent the static closed orbits. The oscillation amplitudes are given by:
 \begin{equation}\label{eq:amplitude}
  A_{(1,2),(x,y)}=\sqrt{2J_{1,2}\beta_{(1,2),(x,y)}(s)},
 \end{equation}
 where $J_{1,2}$ are the linear action-variables associated with the initial excitation, which are global beam parameters consistent across all BPMs, $\beta_{(1,2),(x,y)}(s)$ are the Twiss functions for modes $1$ or $2$ at the location of $s$ in the horizontal $x$ or vertical $y$ planes, respectively. In Eq.~\eqref{eq:coupledTbT}, the radiation damping and beam decoherence are ignored. A generalized Courant-Snyder parameterization developed by Ripken et al.,~\cite{Borchardt1988, Willeke1989, Lebedev2010}, is adopted in Eq.~\eqref{eq:coupledTbT}. Alternative parameterization methods were available, such as ~\cite{Edwards1973,Sagan1999,Luo2004,Wolski2006}, but were not applied in this study.

 In our studies, we employed the harmonic analysis method~\cite{Borer1992} to determine the Twiss $\beta$ functions. Utilizing the spectrum analysis techniques, such as the Fast Fourier Transformation (FFT) or the Numerical Analysis of Fundamental Frequencies (NAFF), the fractional tune $\nu_{1,2}$ can be obtained first. The amplitudes of two harmonics (modes) in each plane were computed using their respective cosine and sine components as follows:
 \begin{equation}\label{eq:harm}
  \begin{array}{ccc}
   C_{(1,2),(x,y)} & = & \sum_{i=1}^N(x,y)_i\cdot\cos(2\pi\nu_{1,2}i)\\
   S_{(1,2),(x,y)} & = & \sum_{i=1}^N(x,y)_i\cdot\sin(2\pi\nu_{1,2}i)
  \end{array}.
 \end{equation}
 By substituting Eqs.~\eqref{eq:coupledTbT} into Eqs.~\eqref{eq:harm}, and averaging over a sufficient number of turns (samples) $N$, the results are:
 \begin{equation}\label{eq:harm1}
  \begin{array}{rrr}
   C_{(1,2),(x,y)} & \approx &
  \frac{N}{2}\sqrt{2J_{1,2}\beta_{(1,2),(x,y)}}\cos(\phi_{(1,2),(x,y)})\\
   S_{(1,2),(x,y)} & \approx &
   \frac{N}{2}\sqrt{2J_{1,2}\beta_{(1,2),(x,y)}}\sin(\phi_{(1,2)(x,y)})
  \end{array}.
 \end{equation}
 The amplitudes $A_{(1,2),(x,y)}$ of the oscillation observed at the BPM are then calculated using:
 \begin{equation}\label{eq:harm2}
  A_{(1,2),(x,y)}=\sqrt{2J_{1,2}\beta_{(1,2),(x,y)}} =
  \frac{2}{N}\sqrt{C_{(1,2),(x,y)}^2+S_{(1,2),(x,y)}^2},
 \end{equation}
 and the initial phases are determined by:
 \begin{equation}\label{eq:harm3}
  \phi_{(1,2),(x,y)}=\arctan\left(\frac{S_{(1,2),(x,y)}}{C_{(1,2),(x,y)}}\right),
 \end{equation}
 where the sign of the phase $\phi$ depends on the signs of $C$ and $S$, ensuring that the phase increases monotonically along the longitudinal direction. This harmonic analysis approach (Eqs.~\eqref{eq:harm}-\eqref{eq:harm3}) can be conducted using other methods such as the NAFF technique~\cite{Laskar1992,Zisopoulos2013}.

 The amplitudes $A$ in Eq.~\eqref{eq:harm2} are intertwined with the global action-variables $J_{1,2}$, and the position-dependent Twiss functions $\beta_{(1,2),(x,y)}(s)$. To determine the local $\beta$-functions, it is essential to calibrate the action-variables, $J$, first. For weakly-coupled linear lattices with an established model, such calibration can usually be achieved by scaling the measured squared amplitudes, $A_j^2$ (with $j$ as the index of BPMs) to match the design lattice model's $\beta_j$ values. This scaling approach, however, may introduce systematic errors in some scenarios. One such example arises when strong coupling is introduced from random unidentified errors with a near resonance tune. In this case, precise lattice models are not always available. To address this challenge, it is necessary to absolutely measure the $\beta$-functions at at least one BPM location, and then extract the global action-variables with Eq.~\eqref{eq:harm2}.

 With the installation of the new BPM pair, the consecutive $4\times(N-1)$ phase space trajectories observed by the BPMs was used to derive the one-turn-matrices $M$ as shown:
 \begin{equation}\label{eq:fitting}
  \begin{pmatrix}
   x_2,x_3,\cdots,x_n \\
   p_{x,2},p_{x,3},\cdots,p_{x,n} \\
   y_2,y_3,\cdots,y_n \\
   p_{y,2},p_{y,3}\cdots,p_{y,n}
  \end{pmatrix}
  =M
  \begin{pmatrix}
   x_1,x_2,\cdots,x_{n-1} \\
   p_{x,1},p_{x,2},\cdots,p_{x,n-1} \\
   y_1,y_2,\cdots,y_{n-1} \\
   p_{y,1},p_{y,2}\cdots,p_{y,n-1}
  \end{pmatrix},
 \end{equation}
 where momenta $p_{x,y}$ were computed with Eq.~\eqref{eq:slope} and $M$ is a $4\times4$ matrix, with N representing the number of turns. From $M$, the Ripken Twiss parameters at the BPM locations can be obtained, enabling the determination of the global action variables $J$ via Eq.~\eqref{eq:harm2}.  These variables are then used to normalize amplitudes measured by other BPMs. This methodology is generally applicable to both weakly and strongly coupled linear lattices, therefore, no detailed accelerator model is needed.

\subsection{Measurement of a weakly coupled lattice}
 First we measured the $\beta$-functions and tunes for a weakly coupled lattice, and compared against the design model. Given a weakly coupled lattice, it is unnecessary to use this method, because the amplitudes can be quite accurately normalized with the existing design model $\beta$-functions as explained previously. However, we can take advantage of this to implement a beam-based calibration for the new BPM pair, similar to the Linear Optics from Closed Orbit (LOCO) technique~\cite{safranek1997}. 
 
\begin{figure}[ht]
    \centering
    \includegraphics[width=\columnwidth]{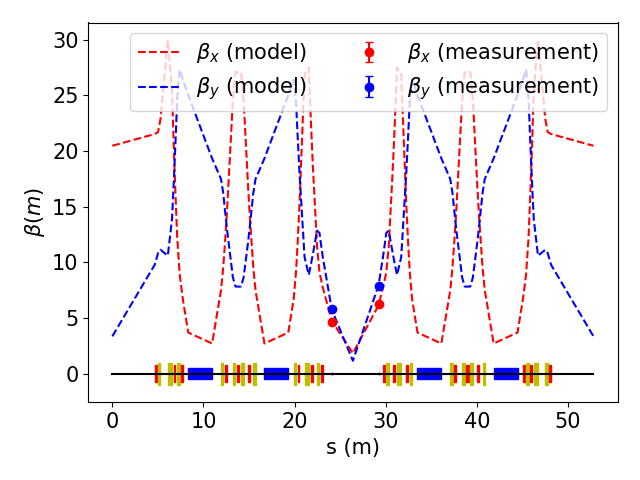}
    \caption{(Colored) Comparison of the measured $\beta$-functions (dots with error-bars) to the design model (dashed lines). The error bars, detailed in Tab.~\ref{tab:weakbeta}, represent the means and standard deviations across 11 snapshots. Here, $s=0$ \si{m} is the location of the injection point.}
    \label{fig:beta} 
 \end{figure}

 A short bunch-train consisting of 10 consecutive buckets and carrying a charge of 1.5 \si{mA} was stored in the storage ring, which was configured with a well-corrected linear lattice. Upon excitation, the 10 bunches began to perform coherent Betatron oscillations. With a T$\times$T data set of 512 turns, the one-turn-matrix $M$ observed at the upstream BPM was determined using Eq.~\eqref{eq:fitting}:
 \begin{equation}{\label{eq:otm}}
  M=\begin{pmatrix*}[r]
 1.370\pm 0.002&   4.608\pm 0.005&  -0.020\pm 0.001&  -0.221\pm 0.004 \\
-0.537\pm 0.001&  -1.075\pm 0.002&  -0.030\pm 0.001&  -0.020\pm 0.001 \\
-0.008\pm 0.002&  -0.171\pm 0.005&   1.852\pm 0.003&   5.751\pm 0.008 \\
-0.012\pm 0.001&   0.025\pm 0.002&  -0.843\pm 0.001&  -2.079\pm 0.003
  \end{pmatrix*}.
 \end{equation}
 The determinant of this matrix, $\det{M}=0.993\pm0.001$, represents the closest measurement to the symplectic condition ($\det{M}=1$) ever achieved at the NSLS-II. From the matrix Eq.~\eqref{eq:otm}, a set of coupled Ripken Twiss functions ($\alpha,\beta,\gamma$) and two eigen-tunes (fractional parts) were extracted and compared against their design values as shown in Tab.~\ref{tab:weakbeta} and illustrated in Fig.~\ref{fig:beta}. The fractional tunes extracted from the T$\times$T data with the NAFF technique closely align with those computed from the OTM. Although the $\beta$-functions, dispersions and phase-advances were simultaneously corrected prior to our measurement, each parameter retains has some residual deviation. Therefore, the tune was not perfectly corrected to the design value. Given the weak coupling in the lattice, the $\beta_{1x,2y}$ functions degenerate into the commonly used, uncoupled $\beta_{x,y}$ functions. Simultaneously, the minor values of $\beta_{1y,2x}$ confirm the coupling is weak. Additionally, two new BPM roll errors were found negligible, and their gains were calibrated as $g_x\approx g_y\approx0.978$, as defined in the LOCO algorithm~\cite{safranek1997}. The BPM calibration was based on comparisons with the scaled action-variables averaged over 180 regular BPMs with a fitted model. Note that those regular BPMs were previously calibrated with the LOCO technique and T$\times$T data~\cite{huang2015}.
 
 \begin{table}
     \caption{Measured and design Ripken $\beta$-functions at the upstream BPM }
     \centering
     \begin{tabular}{|c|c|c|c|}
        \hline
        name (unit) & mean & std. & design\\ 
        \hline\hline
        $\beta_{1,x}$ (\si{m}) & 4.659 & $\pm$0.005 & 4.761\\ 
        $\beta_{2,y}$ (\si{m}) & 5.797 & $\pm$0.008 & 5.765\\ 
        $\beta_{1,y}$ (\si{m}) & 0.007  & $\pm$0.0003 & 0\\ 
        $\beta_{2,x}$ (\si{m}) & 0.004  & $\pm$0.0001 & 0\\
        $\alpha_{1,x}$ & 1.234 & $\pm$0.0017 & 1.258 \\
        $\alpha_{2,y}$ & 1.984 & $\pm$0.0031 & 1.981 \\
        $\alpha_{1,y}$ & 0.001 & $\pm7\times10^{-5}$ & 0\\
        $\alpha_{2,x}$ & 0.003 & $\pm1\times10^{-4}$ & 0 \\
        $\nu_{1,OTM}$ & 0.2264  & $\pm8\times10^{-5}$ & 0.2213\\
        $\nu_{2,OTM}$ & 0.2682  & $\pm5\times10^{-5}$ & 0.2598\\
        $\nu_{1,NAFF}$ & 0.2264  & $\pm8\times10^{-5}$ & 0.2213\\
        $\nu_{2,NAFF}$ & 0.2682  & $\pm4\times10^{-5}$ & 0.2598\\
        \hline
     \end{tabular}
     \label{tab:weakbeta}
 \end{table}

\begin{figure}[ht]
    \centering
    \includegraphics[width=\columnwidth]{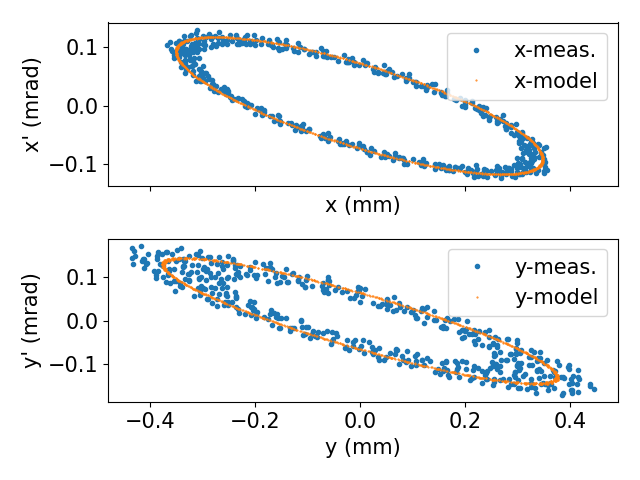}
    \caption{(Colored) Comparison of the measured 512 turns Poincar\'e maps (blue dots) and the design model (oranges dots) as observed by the upstream BPM.}
    \label{fig:phaseSpaceX}
 \end{figure}

 The phase space trajectories observed by the upstream BPM are illustrated in Fig.~\ref{fig:phaseSpaceX}. These trajectories are plotted alongside those predicted by the design model comparative analysis. Notably, in the vertical plane, obvious smearing was observed. Part of the smear might possibly be attributed to not-well-corrected non-linearity of the lattice because of lack of independent tunability of sextupoles powered in series~\cite{li2024}. In the presence of non-linearity, to accurately extract the linear OTM, a sufficient number of data points is needed. In our studies, we generally use 512 data points to mitigate the nonlinear effect before Betatron oscillation is damped by the radiation loss (the damping time: 20,800 turns~\cite{Dierker2007}) and beam decoherence. 

 Minor deviations from the symplectic condition in $M$ could result from factors such as the combination of BPM imperfections, radiation damping, decoherence. A process known as ``symplectification" was applied to the matrix. In our case, the Hamilton-Cayley theorem~\cite{Chao2002} was applied to approximate $M$ to a linear Lie generator, $f_2=C_{abcd}x^ap_x^by^cp_y^d$. Here the indices $a,b,c,d$ are non-negative integers with a sum $a+b+c+d=2$, and $C_{abcd}$ are the coefficients of monomials. Based on the Lie generator $f_2$, another set of Twiss parameters and tunes were computed. The results, however, were nearly identical to those derived directly with $M$. Consequently, we concluded that the symplectification process could be omitted.
 
\begin{figure}[ht]
    \centering
    \includegraphics[width=\columnwidth]{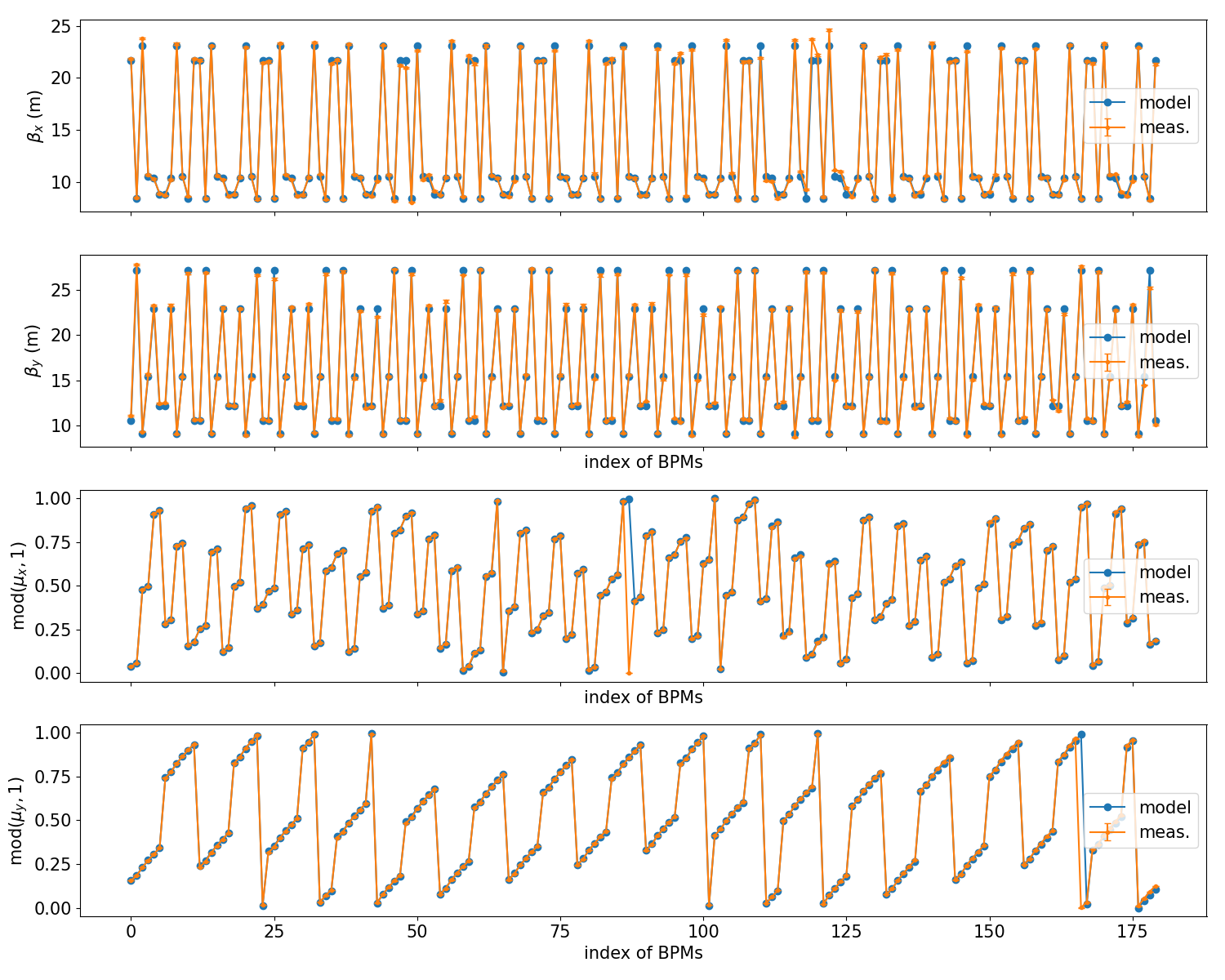}
    \caption{Comparison of the measured $\beta$-functions and phases (orange lines with error-bars) against the design model (blue dotted-lines.), The phases were normalized with $2\pi$ and then displayed with respect to a modulus of 1.}
    \label{fig:weak_couple_beta}
 \end{figure}

 The action-variables $J_{1,2}$ were determined using Eq.~\eqref{eq:harm2} as follows:
 \begin{equation}
    J_1 = 0.0117\pm1.8\times10^{-5}\,\si{m},\; J_2 = 0.0104\pm5.7\times10^{-5}\,\si{m}.
 \end{equation}
 Since the action-variable seen by all BPMs remains consistent, it was used to normalize the amplitudes to determine the $\beta$-functions across the entire ring  as shown in Fig.~\ref{fig:weak_couple_beta}. It is important to note that the measurement of the Betatron phase, as demonstrated in the two bottom subplots, using Eq.~\eqref{eq:harm3}, neither needs a design model, nor depends on the BPM gains. However, a precise alignment of all BPM data to the same turn and/or bunch is essential to measure their phases correctly.

\subsection{Measurement of a strongly coupled lattice}
 In this section, we applied our method to a strongly coupled linear lattice in which a lattice model is unavailable. Initially, all skew quadrupole correctors were turned off and the tune was adjusted such that it was near the difference resonance $\nu_1-\nu_2=17$. In this configuration, linear coupling mainly arises from random and unidentified quadrupole tilts and vertical orbit displacements through sextupoles, and no precise lattice model is available.  Since the beam orbit was already aligned to the centers of the magnets using a beam-based alignment technique~\cite{portmann1995}, the contribution from the quadrupole tilts is considered dominant as explained in ref.~\cite{Li2022}. Therefore, by analyzing the quadrupole tilts with measured lattice functions, we can estimate the overall misalignment of magnets and compare it with survey results. As mentioned previously, the BPM gains and roll errors were pre-calibrated and then excluded from the following coupling measurements.

\begin{figure}[ht]
     \centering
     \includegraphics[width=\columnwidth]{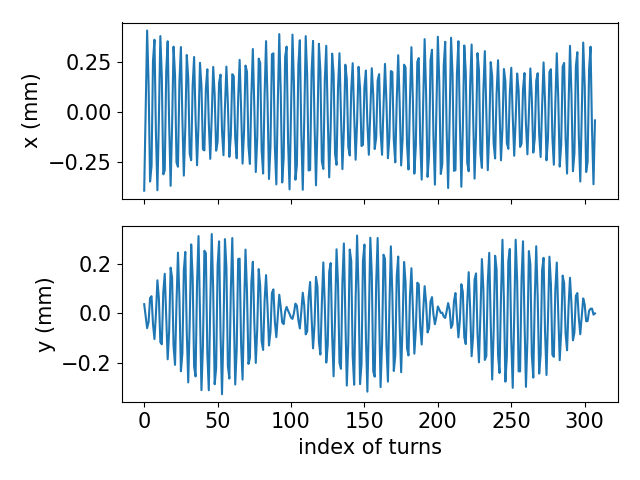}
     \caption{An out-of-phase beat pattern as seen from the horizontal and vertical T$\times$T data under strong linear coupling.}
     \label{fig:strongCoupleTxT}
 \end{figure}

\begin{figure}[ht]
    \centering
    \includegraphics[width=\columnwidth]{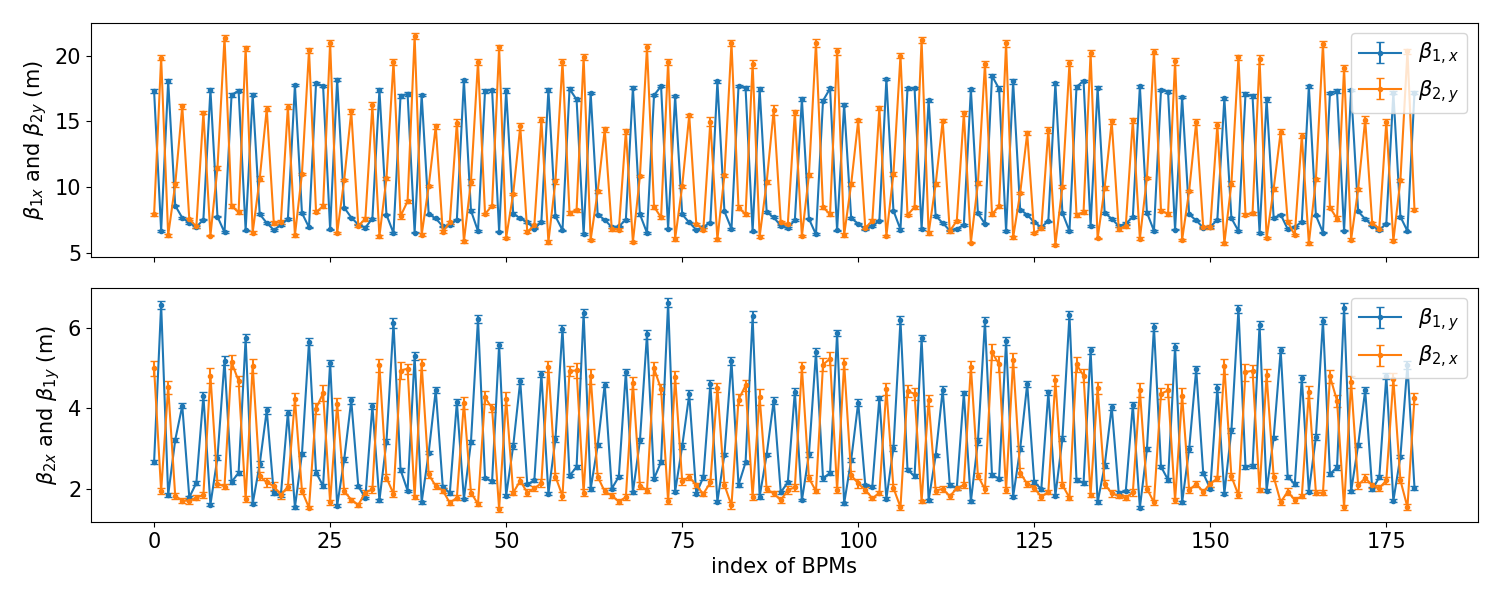}
    \caption{(Colored) Coupled Ripken $\beta$-functions observed by BPMs in a strongly coupled linear lattice, with a near resonance tune $\nu_{1,2}=33.222/16.212$.}
    \label{fig:strong_couple_beta}
 \end{figure}

 From the T$\times$T data collected at 182 BPMs, a distinct beat pattern, due to a strong linear coupling, can be observed as illustrated in Fig.~\ref{fig:strongCoupleTxT}. With the absolute calibrated $\beta$-functions at the BPM pair, as explained in the previous section, the four coupled $\beta$-functions seen by the upstream BPM are: $\beta_{1x}=3.765\pm0.017\,\si{m}$, $\beta_{2y}=3.984\pm0.035\,\si{m}$, $\beta_{1y}=0.926\pm0.016\,\si{m}$, $\beta_{2x}=0.948\pm0.028\,\si{m}$. The two global action-variables can then be extracted as: $J_1= 0.0113\pm 1.1\times10^{-4}\,\si{m},\;J_2= 0.0028\pm 1.4\times10^{-4}\,\si{m}$. The coupled Twiss functions seen by the other 180 BPMs were determined by scaling their amplitudes $A_{(1,2),(x,y)}$ with the actions $J_{1,2}$. The measured lattice functions are illustrated in Fig~\ref{fig:strong_couple_beta}.
 
\begin{figure}[ht]
     \centering
     \includegraphics[width=\columnwidth]{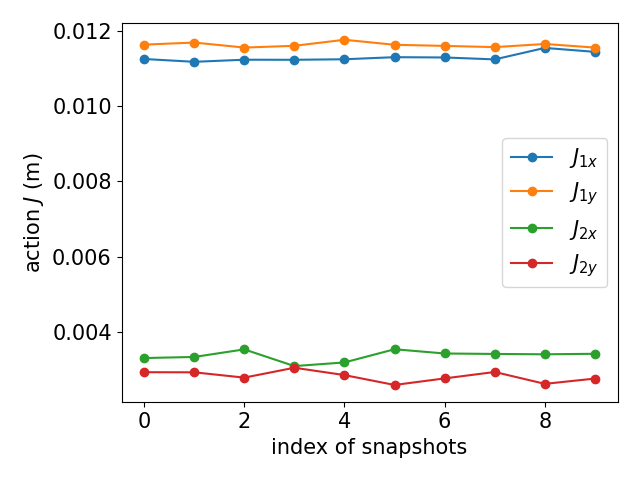}
     \caption{Self-consistency check of linear action-variables measured in two planes, respectively. Each snapshot represents an independent measurement.}
     \label{fig:actionCheck}
 \end{figure}

 In a strongly coupled lattice, two linear action-variables projected onto two planes can be calibrated independently, providing a method for self-consistency checks. For example, $J_1$ can be calibrated either in the horizontal or vertical plane as follows:
 \begin{equation}
      J_1=\frac{A_x^2}{2\beta_{1,x}}=\frac{A_y^2}{2\beta_{1,y}},
 \end{equation}
 and similarly for $J_2$. The self-consistency check is illustrated in Fig.~\ref{fig:actionCheck}.
 In Fig.~\ref{fig:actionCheck}, the action $J_1$ is approximately 3 times larger than $J_2$, which results in a better signal-noise-ratio, thereby enhancing the self-consistency measurements. 
 
\begin{figure}[ht]
     \centering
     \includegraphics[width=\columnwidth]{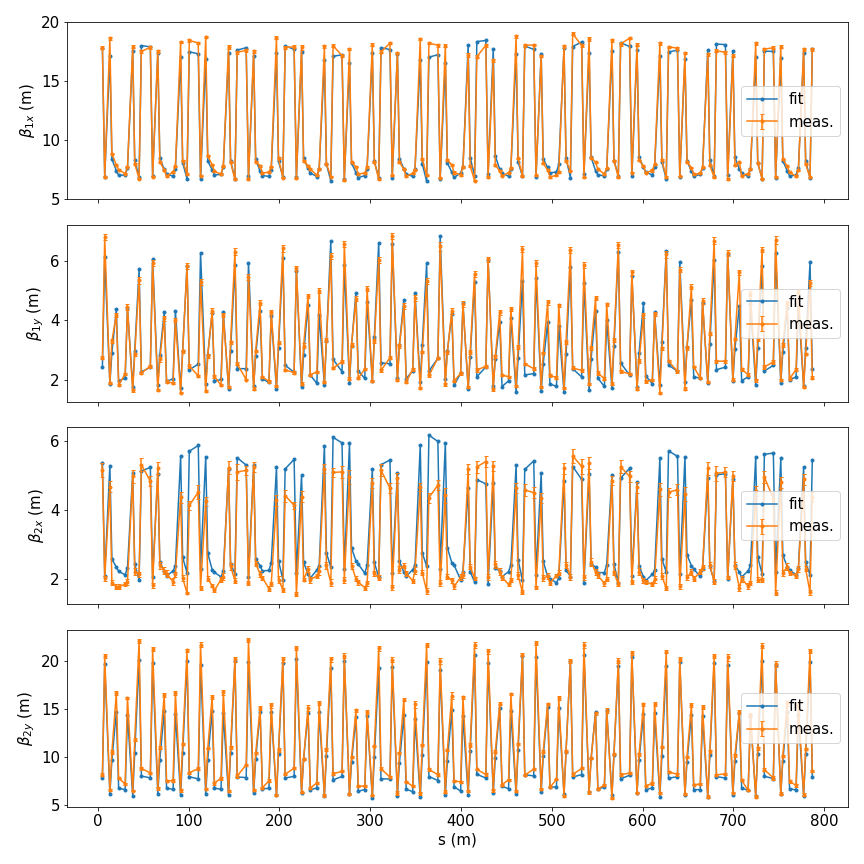}
     \caption{Fitting of a linear lattice model based on the measured $\beta$-functions shown in Fig.~\ref{fig:strong_couple_beta}.}
     \label{fig:strongCoupleFit}
 \end{figure}

 With the measured $\beta$-functions, we fitted a linear lattice model by adjusting the tilt of all 300 quadrupoles. The model was constructed by using an ``approximate" Jacobian matrix:
 \begin{equation}\label{eq:betaJac}
    \textrm{J}_{\beta} =
    \left[
      \frac{\partial\beta_{(1,2),(x,y)}}{\partial\theta_1},
      \cdots,
      \frac{\partial\beta_{(1,2),(x,y)}}{\partial\theta_n}\right],
 \end{equation}
 here, $\beta$-functions are at the locations of BPMs, and $\theta_{1,2,\cdots,n}$ are the $n$ quadrupole tilts. Lacking a precise linear lattice model, this Jacobian matrix of dimension $720\times300$ for $180\times4$ $\beta$-functions, and seen by 180 BPMs against 300 quadrupole tilts, was computed  with a ``similar" near-resonance coupled lattice. This similar lattice, which has the same tunes as the measured one, only represents approximately the response of the $\beta$-function to the quadrupole tilt. The near-resonance tune allows minor changes in quadrupole tilt to generate an obvious coupled lattice change. The approximate Jacobian matrix was updated iteratively through the fitting. After several iterations, a linear lattice model which was reasonably consistent with the measurements, was obtained and illustrated in Fig.~\ref{fig:strongCoupleFit}. Convergence was generally achieved with the exception of $\beta_{2x}$. 
  
 Some possible explanations for the imperfect convergence (particularly in $\beta_{2x}$) may include: (1) some coupling contributions by the orbit's vertical offsets through sextupoles, indicating that merely fitting quadrupole tilts might be insufficient; (2) the large number of fitting parameters (quadrupole tilts in this case) relative to the objectives ($4\times$180 $\beta$-functions) may prevent optimal fitting of our model. Nevertheless, based on the fitted model, we estimated the quadrupole tilt's root-mean-square (r.m.s.) to be around 400 \si{\mu}rad. As previously mentioned, part of the coupling originates from the orbit offsets in sextupoles. Therefore, this suggests that the actual quadrupole tilts at the NSLS-II ring should be better than estimated. The specification of magnet tilts for installation is 200 \si{\mu}rad~\cite{Dierker2007}. After the 10-years of operation and despite potential other issues, such as settling of the floor and other infrastructure components, the magnet alignment still maintains good accuracy. This is the first time we estimated our quadrupole roll errors based on a beam test.

\begin{figure}[ht]
     \centering
     \includegraphics[width=\columnwidth]{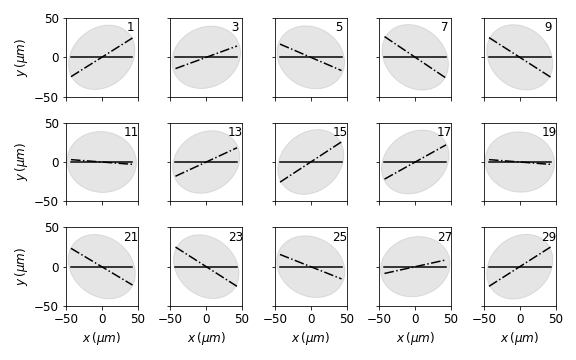}
     \caption{Transverse beam profiles at 15 short straight centers (indexed at the upright corners of subplots) by incorporating the fitted quadrupole tilts into the lattice model. Dashed lines indicate the tilts of beam profiles relative to the horizontal plane.}
     \label{fig:strongCoupleBeamSize}
 \end{figure}

 Using the fitted model, the equilibrium eigen-emittances computed using the SLIM technique~\cite{Chao2002,Li2022} were $\epsilon_{1,2}=1.6/0.5\,\si{nm}$. This also allowed for estimation of local beam sizes throughout the entire storage ring. Fig.~\ref{fig:strongCoupleBeamSize} illustrates the transverse beam profiles at 15 short, straight centers. The tilts of these profiles, represented by  $\tan(2\theta_{xy})=\frac{2\left<xy\right>}{\left<xx\right>-\left<yy\right>}$ were marked with dashed lines. Here $\left<xy\right>$, $\left<xx\right>$ and $\left<xy\right>$ are the second moments of beam distributions. Variations in beam size, and the presence of random tilts $\theta_{xy}$ result from the randomness in quadrupole tilt distributions. The beam size information can be used for computation of some critical beam parameters, such as beam lifetime and undulator brightness.
  
\subsection{Comparison with a model-dependent measurement result}
\begin{figure}[ht]
     \centering
     \includegraphics[width=\columnwidth]{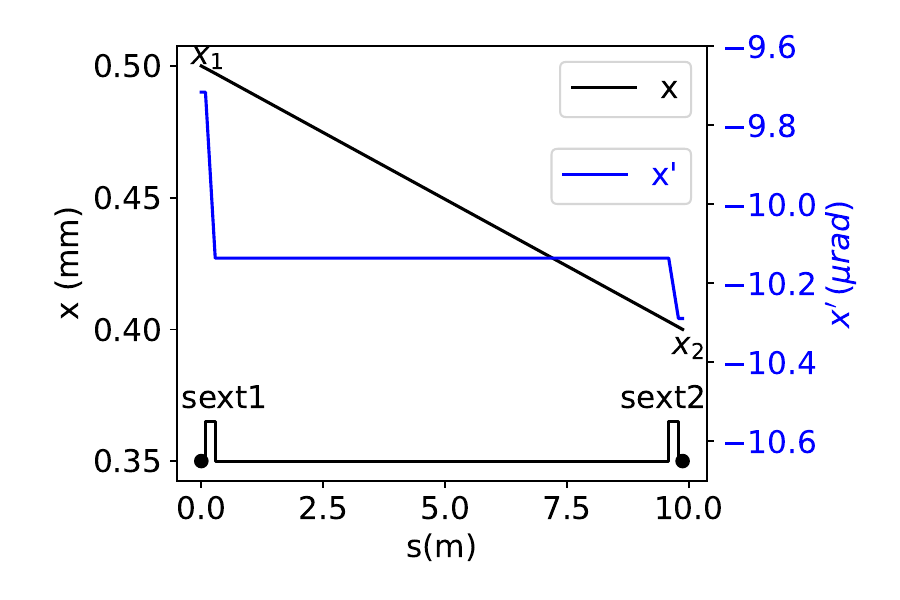}
     \caption{Reconstruction of the proximal trajectory within an NSLS-II straight section. The black dots represent the readings from a pair of BPMs separated by two sextupoles. The black line depicts the reconstructed trajectory in the horizontal plane, while the blue line represents the slope $x^{\prime}$, which experiences jumps while passing through sextupoles. The kicks received from the sextupoles are sub-\si{\mu}rad, therefore, the trajectory appears almost linear between the BPM pair.}
     \label{fig:nsls2ls}
 \end{figure}

 In this section, another model-dependent measurement of OTMs was implemented across all 30 straight sections of the NSLS-II. Each section features a pair of BPMs, flanking two sextupoles. In this configuration, trajectories in the phase space can be reconstructed by incorporating the actual settings of the magnets as illustrated in Fig.~\ref{fig:nsls2ls}. By using the known magnet settings and the readings of the two BPMs, the trajectories can be numerically resolved through tracking simulations. 
 
 When nonlinear magnets such as sextupoles are positioned between two BPMs, and only BPM readings are available, multiple trajectories can theoretically exist. In practice, however, only the proximal trajectory is real and viable, as others would extend beyond the dynamic aperture, and thus are not feasible. Nevertheless, this method still relies on certain assumptions, including a pre-established lattice model and known magnetic fields. 
 
\begin{figure}[ht]
     \centering
     \includegraphics[width=\columnwidth]{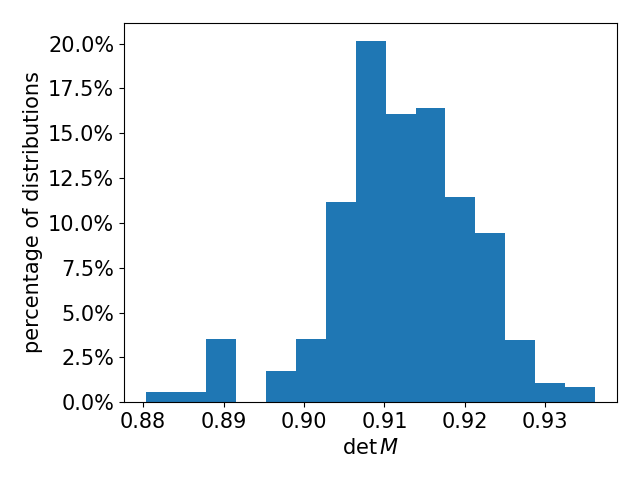}
     \caption{Distribution of $\det{M}$'s measured at NSLS-II straight sections where sextupoles intervene. These $\det{M}$'s are not only significantly smaller than unity, but they also exhibit large fluctuations. For context, the matrices derived from the new BPMs show $\det{M}=0.9933\pm0.0005$, illustrating better symplectic integrity.}
     \label{fig:detM}
 \end{figure}

 Using the same dataset as the previous section, the OTMs were reconstructed at these BPM pairs. The symplectic quality of matrices, assessed by evaluating their determinants, was found to be about 10\% ($\left<\det{M}\right>=0.91$) lower than those derived from new BPMs ($\det{M}=0.99$) as depicted in Fig.~\ref{fig:detM}. The errors of $\beta$-functions and tunes derived from these non-symplectic matrices can propagate and influence the characterization of the full storage ring lattice.
 
\subsection{Observation of non-linearity}
\begin{figure}
     \centering
     \includegraphics[width=0.8\columnwidth]{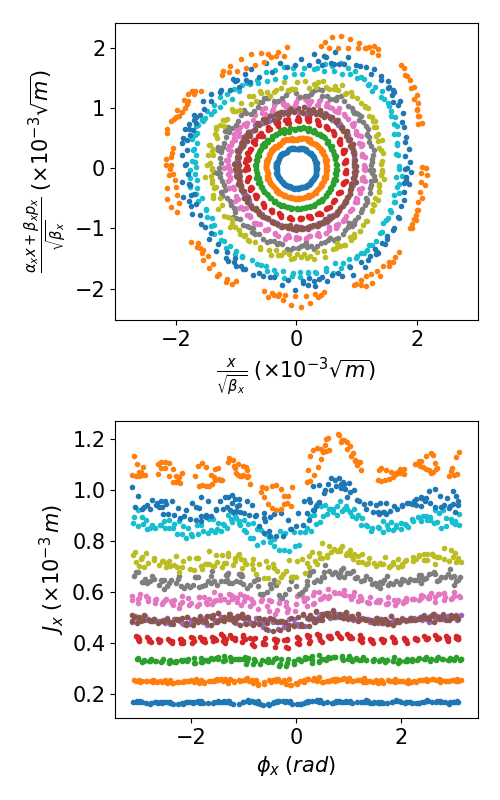}
     \caption{Visualization of lattice non-linearity in normalized horizontal coordinates. Top: phase space trajectories in normalized Courant-Synder coordinates. Bottom: Modulation of linear actions with their conjugate angle-variables.}
     \label{fig:phase_DA}
 \end{figure}

 With the new BPM pair, the non-linearity of the lattice was observed more accurately by projecting it into the normalized phase-space and action-angle variables. Initially, the beam was excited with sufficiently small kicks to keep its motion approximately linear, allowing for precise measurement of the linear Twiss functions ($\beta$'s and $\alpha$'s) as shown in Tab.~\ref{tab:weakbeta}. Subsequently, the beam was excited with gradually increasing kicks until non-linearity became obvious. Notably, the new BPM pair allows for accurate measurements of not only the $\beta$-functions, but $\alpha$-functions as well. Through canonical transformations, we derived the $s$-independent normalized Courant-Synder coordinates~\cite{courant1958} $\bar{x}=\frac{x}{\sqrt{\beta_x}},\,\bar{p}_x=\frac{\alpha_x x+\beta_xp_x}{\sqrt{\beta_x}}$, and action-angle variables $J_x=\frac{1}{2}\sqrt{\bar{x}^2+\bar{p}_x^2},\,\phi_x=\arctan\left(\frac{\bar{p}_x}{\bar{x}}\right)$ as illustrated in Fig.~\ref{fig:phase_DA}. The deformation of trajectories and the modulation of action-variables with their conjugate angles served as measures of non-linearity, which can potentially be used as objectives for online lattice optimizations~\cite{li2018}.
 
\section{\label{sect:BxB}Bunch-by-bunch Twiss functions measurements}
\begin{figure}[ht]
     \centering
     \includegraphics[width=\columnwidth]{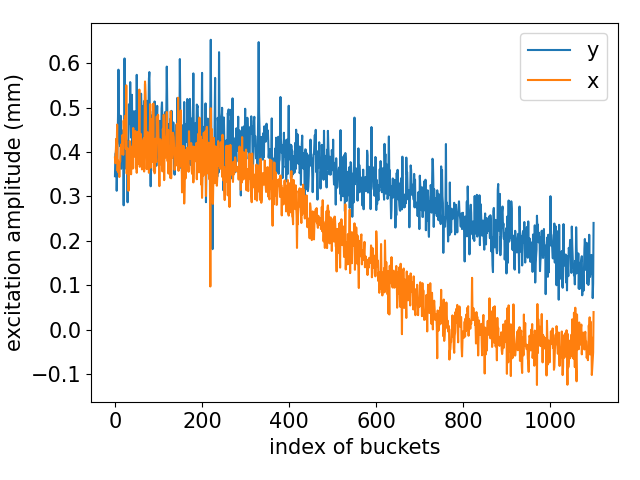}
     \caption{Initial excitation amplitudes along the bunch-train in the horizontal and vertical planes. This profile correlates with the pinger's wave-forms.}
     \label{fig:pingerWaveForm}
 \end{figure}

 As mentioned previously in Sect.~\ref{sect:BPM}, this pair of BPMs can resolve individual bunch Poincar\'e map with a 2 \si{ns} bunch separation. This section describes preliminary beam optics measurements that leverage this capability. For this part of the investigation, 25 \si{mA} beam was stored in the storage ring, uniformly distributed in a bunch-train of 1200-buckets. As the NSLS-II storage ring has 1320-buckets, this leaves a gap of 120 empty buckets. The beam was excited using pingers and a B$\times$B dataset for all 1200-filled buckets was recorded over 1000 turns. By examining their first turn data after pinger-excitation, we observed the profile of initial excitation amplitudes along the bunch-train, which correlates well with the pingers' wave-forms as illustrated in Fig.~\ref{fig:pingerWaveForm}. The pingers' peaks were aligned to reference bucket 1, i.e., the head of bunch-train. The flat tops of the wave-forms were about 150 buckets (300 \si{ns}) wide. For bunch trains exceeding this length, the T$\times$T data averaged over the bunch train, post-excitation, is unsuitable for lattice measurements due to incoherent bunch motions. Currently, the regular NSLS-II BPMs are limited to the T$\times$T resolution; therefore, we rely on short bunch-train's T$\times$T data to characterize the linear lattice. It was noted that the current B$\times$B resolution ($37$--$63\,\si{\mu m}$, depending on the bucket filling pattern and beam current) is approximately 2 to 3 times less precise than the regular BPM's T$\times$T resolution, as observed by the fluctuations shown in Fig.~\ref{fig:pingerWaveForm}. Despite this, the B$\times$B capability still provides a potent diagnostic tool for investigating beam dynamics within bunch trains, as explained below.
 
\begin{figure}[ht]
     \centering
     \includegraphics[width=\columnwidth]{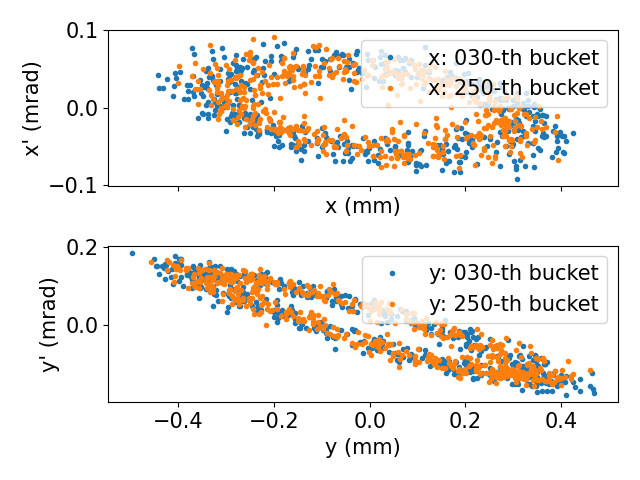}
     \caption{Reconstructed phase space trajectories of the $30^{th}$ and $250^{th}$ bunches at the upstream BPM. The $250^{th}$ bunch's amplitudes (measured with the greater B$\times$B capabilities) are slightly smaller than the $30^{th}$ bunch's due to its smaller initial excitation. From the B$\times$B data, each individual bunch's Twiss functions were extracted.}
     \label{fig:bxbPhaseSpace}
 \end{figure}

\begin{figure}[ht]
     \centering
     \includegraphics[width=\columnwidth]{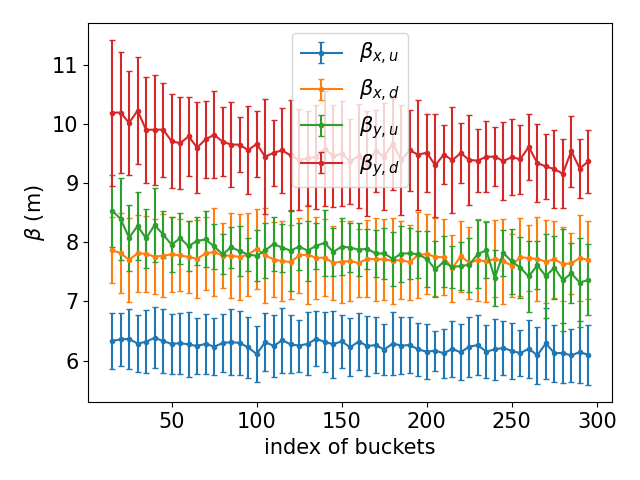}
     \caption{Variations in $\beta$-functions along a long bunch train, as seen by the BPM pair. Here, the subscript ``u" represents the upstream BPM and ``d" represents the downstream BPM. The vertical $\beta$-functions have a visible variation, which might indicate larger vertical wake fields.}
     \label{fig:betaLongTrain}
 \end{figure}

 Using the B$\times$B data, we reconstructed the Poincar\'e maps for each bunch as shown in Fig.~\ref{fig:bxbPhaseSpace}. Following reconstruction, their OTMs were fitted to compute the Twiss parameter along the bunch train, depicted in Fig.~\ref{fig:betaLongTrain}. Despite the relatively low stored beam current of 25 \si{mA}, the linear Twiss functions have notable variation along the bunch-train, particularly in the vertical plane. This variation suggests the existence of a larger vertical than horizontal impedance. Considering the small vertical aperture vacuum chambers used in dipole and ID sections, this result is consistent with the existing NSLS-II infrastructure. 
 
 Should BPMs with the same B$\times$B capabilities be installed throughout the entire storage ring, it would be feasible to re-construct the entire linear lattice model for each individual bunch. Estimation of the quadrupolar wake field distributions would also be possible. Additionally, another potential application of the greater B$\times$B resolution would be visualization of fast oscillating perturbations, such as the residual motion resulting from imperfections in the injection transient.
 
\section{\label{sect:summary}Summary}
 This paper summarizes recent lattice characterization efforts conducted at the NSLS-II ring, facilitated by the installation of a new BPM pair. With the installation location being free of magnetic fields, the BPM pair is particularly useful when precise lattice models are unavailable, or when the online machine significantly deviates from its design model. Through analysis of a strongly coupled lattice, the quadrupole tilt errors were estimated. Additionally, the unprecedented B$\times$B resolution of the BPM pair allowed for observation of the linear lattice variation along a long bunch-train. It is worth noting, however, with the greater instrumentation capabilities provided by the BPMs discussed here, that more extensive studies are currently taking place, but not covered in this manuscript. Some of these studies include investigation into areas such as: collective beam effects with different filling pattern configurations, as well as nonlinear dynamics optimization through Poincar\'e maps analyses for large amplitude beam motions.

\section*{Acknowledgements}
 We would like to extend our thanks to numerous NSLS-II colleagues for their invaluable support, particularly Q. Shen, T. Shaftan, M. Capotosto, S. Wilkins, S. Campbell, C. Yu and the accelerator coordination and operation groups. Their contributions to the project management, hardware setup and beam studies have been essential. This research was supported and funded by the National Synchrotron Light Source-II's Facility Improvement Projects (FIP), and utilized resources of the NSLS-II, a U.S. Department of Energy (DOE) Office of Science User Facility operated for the DOE Office of Science by Brookhaven National Laboratory under Contract No. DE-SC0012704.

\bibliography{ref.bib}

\end{document}